\documentclass[12pt]{article}
\usepackage{graphicx}
\newcommand{\lesssim}{\raise.3ex\hbox{$<$\kern-.75em\lower1ex\hbox{$\sim$}}}
\newcommand{\grtssim}{\raise.3ex\hbox{$>$\kern-.75em\lower1ex\hbox{$\sim$}}}

\title{ Hot Neutron and Quark Star Evolution
\thanks{Lectures
delivered at the Helmholtz International Summer School and
Workshop on {\sl Hot points in Astrophysics and Cosmology}, JINR,
Dubna, Russia, August 2 - 13, 2004.}}

\author{Hovik Grigorian\thanks{On leave from
Yerevan State University, Alex Manoogian Str. 1,
375025 Yerevan, Armenia.}\\[3mm]
Institut f\"ur Physik , Universit\"at Rostock, \\ D-18051 Rostock,
Germany}

\begin{document}
\maketitle

\begin{abstract}
The physics of compact objects is one of the very actively
developing branches of theoretical investigations, since the
careful analysis of different models for the internal structure of
these objects, their evolutionary characteristics and the
comparison with modern observational data give access to
discriminate among various speculations about the state of matter
under extreme conditions.

  The lecture provides an overview to the problem of neutron star
cooling evolution. We discuss the scheme and necessary inputs for
neutron star cooling simulations from the background of a
microscopic modeling in order to present the surface temperature -
age characteristics of hybrid neutron stars including the
determination of main regulators of the cooling process, the
question of neutrino production and diffusion.
\end{abstract}



\newpage
\section{Introduction}

The interiors of compact stars are considered as systems where
high-density phases of strongly interacting matter do occur in
nature, see Shapiro and Teukolsky  \cite{ST}, Glendenning
\cite{Glendenning:wn} and Weber \cite{W99} for textbooks. The
consequences of different phase transition scenarios for the
cooling behaviour of compact stars have been reviewed recently in
comparison with existing X-ray data \cite{Page:2004fy}.

The Einstein Observatory was the first that started the
experimental study of surface temperatures of isolated neutron
stars (NS). Upper limits for some sources have been found. Then
ROSAT offered first detections of surface temperatures. Next
$X$-ray data came from Chandra and XMM/Newton. Appropriate
references to the modern data can be found in recent works by
\cite{Tsuruta:2002ey,Tsuruta:2004ue,Kaminker:2001eu,Yakovlev:2003qy},
devoted to the analysis of the new data. More  upper limits and
detections are expected from satellites planned to be sent in the
nearest future. In general, the data can be separated in three
groups. Some data show very {\em{``slow cooling''}} of objects,
other demonstrate an {\em{``intermediate cooling''}} and some show
very {\em{``rapid cooling''}}. Now we are at the position to
carefully compare the data with existing cooling calculations.

The ``standard'' scenario of neutron star
cooling is based on the main process responsible for the cooling,
which is the modified Urca process (MU) $nn\rightarrow
npe\bar{\nu}$ calculated using the free one pion exchange between
nucleons, see \cite{FM79}. However, this scenario explains only
the group of slow cooling data. To explain a group of rapid
cooling data ``standard'' scenario was supplemented by one of the
so called ``exotic'' processes either with pion condensate, or
with kaon condensate, or with hyperons, or involving the direct
Urca (DU) reactions, see \cite{T79,ST} and refs therein. All these
processes may occur only for the density higher than a critical
density, $(2\div 6)~n_0$, depending on the model, where $n_0$ is
the nuclear saturation density. An other alternative to ''exotic''
processes is the DU process on quarks related to the phase
transition to quark matter.

  Particularly the studies of cooling evolution of compact objects can
give an opportunity for understanding of properties of cold quark
gluon plasma.
 In dense quark matter at temperatures below $\sim 50$ MeV, due
to attractive interaction channels, the Cooper pairing instability
is expected to occur which should lead to a variety of possible
quark pair condensates corresponding to color superconductivity
(CSC) phases, see \cite{Alford:2001dt} for a review.

Since it is difficult to provide low enough temperatures for CSC
phases
in heavy-ion collisions, only precursor phenomena 
\cite{Kitazawa:2001ft,Voskresensky:2004jp} are expected under
these conditions.

CSC phases may occur in neutron star interiors \cite{nsi} and
could manifest themselves, e.g., in the cooling behavior
\cite{Blaschke:1999qx,Page:2000wt,Blaschke:2000dy,
Grigorian:2004jq}.

However, the domain of the QCD phase diagram
where neutron star conditions are met is not yet accessible to
Lattice QCD studies and theoretical approaches have to rely on
non-perturbative QCD modeling. The class of models closest to QCD
are Dyson-Schwinger equation (DSE) approaches which have been
extended recently to finite temperatures and
densities \cite{Bender:1996bm
}. Within simple, infrared-dominant DSE models early studies of
quark stars \cite{Blaschke:1998hy} and  diquark condensation
\cite{Bloch:1999vk} have been performed.


    Estimates of the cooling evolution have been performed
\cite{Blaschke:1999qx} for a self-bound isothermal quark core
neutron star (QCNS) which has a crust but no hadron shell, and for
a quark star (QS) which has neither crust nor hadron shell. It has
been shown there in the case of the 2SC (3SC) phase of QCNS that
the consequences of the occurrence of gaps for the cooling curves
are similar to the case of usual hadronic neutron stars (enhanced
cooling). However, for the CFL case it has been shown that the
cooling is extremely fast since the drop in the specific heat of
superconducting quark matter dominates over the reduction of the
neutrino emissivity. As has been pointed out there, the abnormal
rate of the temperature drop is the consequence of the
approximation of homogeneous temperature profiles the
applicability of which should be limited by the heat transport
effects. Page et al. (2000)\cite{Page:2000wt} estimated the
cooling of hybrid neutron stars (HNS) where heat transport effects
within the superconducting quark core have been disregarded.
Neutrino mean free path in color superconducting quark matter have
been discussed in \cite{cr00} where a short period of cooling
delay at the onset of color superconductivity for a QS has been
conjectured in accordance with the estimates of
\cite{Blaschke:1999qx} in the CFL case for small gaps.

A completely new situation might arise if the scenarios suggested
for (color) superconductivity \cite{arw98,r+98} besides of
bigger pairing gaps ($\Delta_q \sim 50 \div 100$ MeV) will allow
also small diquark pairing gaps ($\Delta_q < 1$ MeV) in quark
matter.

 The questions which should be considered within these
models are the following: (i) Is strange quark matter relevant for
structure and evolution of compact stars? (ii) Are stable hybrid
stars with quark matter interior possible? (iii) What can we learn
about possible CSC phases from neutron star cooling? Further on in
this lectures we discuss the scheme and the results of realization
of the these points in relation with the cooling evolution of
compact objects.

In the consideration of the scenario for the thermal evolution of
NS and HNS we include the heat transport in both the quark and the
hadronic matter. We will demonstrate the influence of the
diquark pairing gaps and the hadronic gaps on the evolution of the
surface temperature.

\begin{figure}[ht]
\begin{center}
\includegraphics[width=0.5\textwidth]{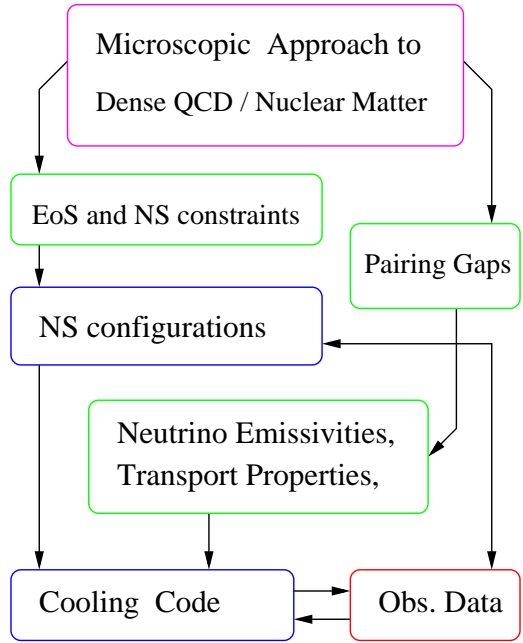}
\caption{The scheme linked problems and tasks, which has to be done
to simulate the cooling of compact objects}
\label{scheme}
\end{center}
\end{figure}

The main strategy of the simulation of the cooling evolution of
compact objects is presented in Fig \ref{scheme}. On the top of
scheme we have the general theoretical background of QCD models as
it has been discussed in the introduction. On the second level of
the scheme we separate two branches one for the structure of the
compact objects and the other for the thermal properties of the
stellar matter. On the bottom of the scheme those two branches are
combined in the code of the cooling simulations project and entail
the comparison of the theoretical and observational results.

\section{EoS of stellar matter and  Compact Star configurations}

\subsection{Hadronic  matter EoS and DU threshold}

For the modeling of the dense hadronic matter different
approaches, like relativistic and non-relativistic
Br\"uckner-Hartree-Fock approach including the three body
correlations
\cite{APR98,Baldo:1997ag,Zuo:2002sf,Heiselberg:1999fe}
 and
different parameterizations of relativistic mean field models, are
discussed successfully in the literature
\cite{Gaitanos:2003zg,DaFuFae04,Typel:2005ba,Kolomeitsev:2002pg}.

For the appropriate cooling simulations we exploit the EoS of
\cite{APR98} (specifically the Argonne $V18+\delta v+UIX^*$
model), which is based on the most recent models for the
nucleon-nucleon interaction including a parameterized
three-body force and relativistic boost corrections. Actually we
adopt a simple analytic parameterization of this model by
Heiselberg and Hjorth-Jensen \cite{Heiselberg:1999fe}, hereafter
HHJ.

The latter uses the compressional part with the compressibility
$K\simeq 240$~MeV, a symmetry energy fitted to the data around
nuclear saturation density and smoothly incorporates causality at
high densities. The density dependence of the symmetry energy is
very important since it determines the value of the threshold
density for the DU process. The HHJ EoS fits the symmetry energy
to the original Argonne $V18+\delta v +UIX^*$ model yielding
$n_c^{\rm DU}\simeq~5.0~n_0$ ($M_c^{\rm DU}\simeq
1.839~M_{\odot}$).

\begin{figure}[ht]
\begin{center}
\includegraphics[width=0.5\textwidth, angle = -90]{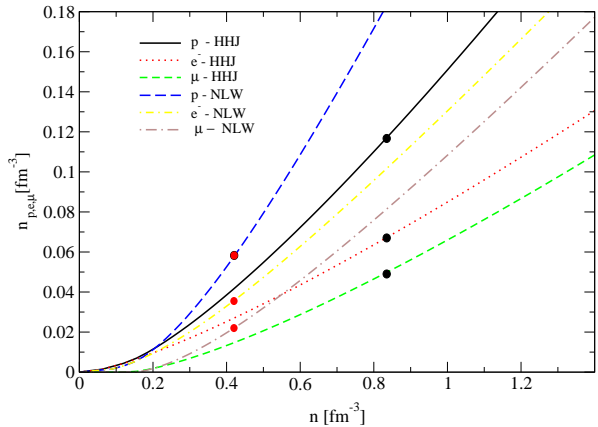}
\caption{Densities of the charged particles as a function of
baryon density for HHJ and NLW model EoS. Possibility of pion
condensation is suppressed.\label{fig3} }
\end{center}
\end{figure}

Fig. \ref{fig3} demonstrates the partial densities of $p$, $e$ and
$\mu^-$ in HHJ and relativistic non-linear Walecka (NLW) model in
the parameterization of \cite{Kolomeitsev:2002pg}, adjusted to the
following bulk parameters of the nuclear matter at saturation:
$n_0 =0.16$~fm$^{-3}$. Anyhow, in the given NLW model the
threshold density for the DU process is $n_c^{\rm
DU}\simeq~2.7~n_0$.

\subsection{Quark matter EoS with 2SC superconductivity}

 The quark matter models~\cite{r+98,alford99,Blaschke:1998md} predict,
that the diquark pairing condensate is possible for temperatures
and densities relevant for compact objects. The order of magnitude
of pairing gaps is 100 MeV and a remarkably rich phase structure
of the matter has been
identified~\cite{Alford:2001dt,Rajagopal:2000wf,Buballa:2003qv,Schmitt:2004et}.

This leads to expectations of observable effects of color
superconducting phases, particularly in the compact star cooling
behaviour~\cite{Blaschke:1999qx,Page:2000wt,Blaschke:2000dy,Grigorian:2004jq,Horvath:1991ms}.
Generally color superconductivity is involved in all aspects of
neutron star studies, such as magnetic field
evolution~\cite{Blaschke:1999fy,Alford:1999pb,Iida:2002ev,Sedrakian:2002ff}
or burst-type
phenomena~\cite{Hong:2001gt,Ouyed:2001cg,Aguilera:2002dh,Ouyed:2005tm}.

The applications of the nonlocal chiral quark model developed in
\cite{Blaschke:2003yn} for the case of neutron star constraints
shows that the relevant CSC phase is a 2SC phase while the
omission of the strange quark flavor is justified by the fact that
chemical potentials in central parts of the stars do barely reach
the threshold value at which the mass gap for strange quarks
breaks down and they appear in the system
\cite{Gocke:2001ri,Blaschke:2005uj}.

It has been shown in that work that the Gaussian formfactor ansatz
 of quark interaction (hereafter we call it SM model) leads to an
early onset of the deconfinement transition and such a model is
therefore suitable to discuss hybrid stars with large quark matter
cores \cite{Grigorian:2003vi}.

\begin{figure}[htb]
  \begin{center}
    \includegraphics[width=0.45\linewidth,angle=-90]{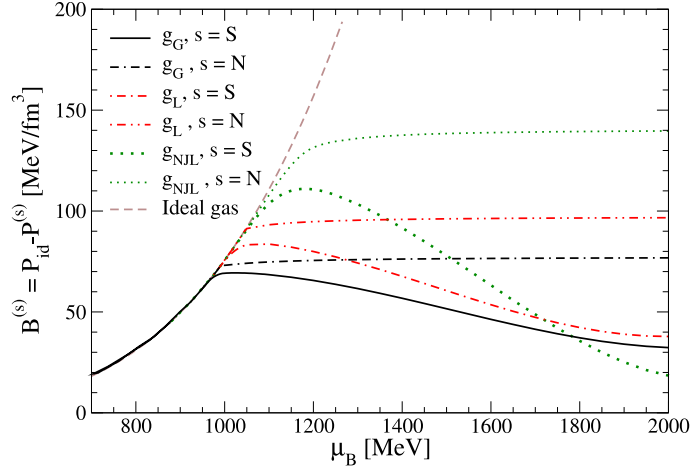}
    \vspace*{0.7cm}
    \caption{Bag pressure for different formfactors of the quark interaction
in dependence on the baryon chemical potential for  $\eta = 0$ and
for  $\eta = 1$. For the latter the superconducting phase is
realized.}
    \label{fig:bagnew}
  \end{center}
\end{figure}

The resulting quark matter EoS within this nonlocal chiral model
\cite{Blaschke:2003yn} can be represented in a form reminiscent of
a bag model
\begin{equation}
\label{press}
  P^{(s)} = P_{id}(\mu_B) - B^{(s)}(\mu_B)~,
\end{equation}
where $P_{id}(\mu_B)$ is the ideal gas pressure of quarks and
$B^{(s)}(\mu_B)$ a {\it density dependent} bag pressure, see Fig.
\ref{fig:bagnew}. The occurrence of diquark condensation depends
on the value of the ratio $\eta=G_2/G_1$ of coupling constants and
the superscript $s\in\{S,N\}$ indicating whether we consider the
matter in the superconducting mixed phase ($\eta=1$) or in the
normal phase ($\eta=0$), respectively.

\subsection{EoS of hybrid star matter}

In some density interval above the onset of the first order phase
transition, there may appear a mixed phase region between the
hadronic (confined quark phase) and deconfined quark phases, see
\cite{G92}. Ref. \cite{G92} disregarded finite size effects, such
as surface tension and charge screening. On the example of the hadron-quark mixed phase Refs \cite{VYT02} demonstrated that
finite size effects might play a crucial role substantially
narrowing the region of the mixed phase or even forbidding its
appearance in a hybrid star configuration. Therefore we omit the
possibility of the hadron-quark mixed phase in our model assuming
that the quark phase arises by the Maxwell construction.

\begin{figure}[hth]
  \begin{center}
    \vspace*{0.7cm}
    \includegraphics[width=0.45\linewidth,angle=-90]{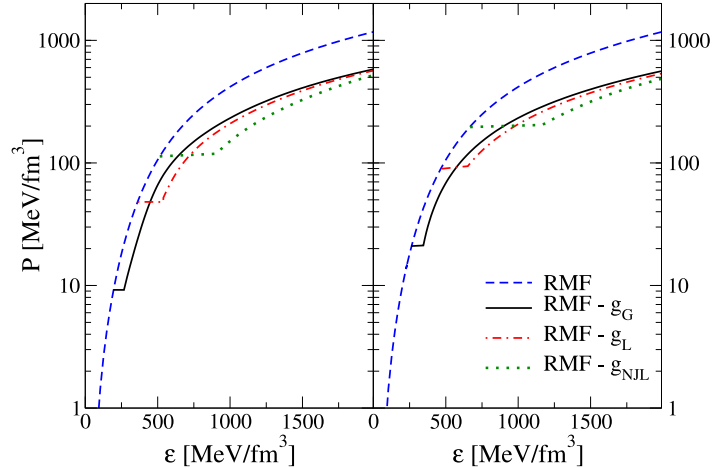}
    \caption{EoS for strongly interacting matter at zero temperature under
compact star constraints for the coupling parameter $\eta=1$ (left
panel) and $\eta=0$ (right panel). Dashed line: relativistic
mean-field model for hadronic matter; solid, dash-dotted and
dotted lines correspond to quark matter with Gaussian, Lorentzian
and NJL formfactor functions, respectively.}
    \label{fig:Eos_2in1}
  \end{center}
\end{figure}

To demonstrate the EoS with quark-hadron phase transition
applicable for hybrid star configurations in the next Section, we show in
Fig. \ref{fig:Eos_2in1} results using the relativistic
mean field (RMF) model of asymmetric nuclear matter including a
non-linear scalar field potential and the $\rho$ meson (nonlinear
Walecka model) ( see \cite{Glendenning:wn}) for the cases $\eta=1$
(left panel) when the quark matter phase is superconducting and
for $\eta=0$ (right panel) when it is normal.

In case the hadonic EoS is chosen to be HHJ model and the quark
one is SM model we found a tiny density jump at the phase-boundary
from $n_c^{\rm hadr}\simeq 0.44~{\rm fm}^{-3}$ to {\bf{$n_c^{\rm
quark}\simeq 0.46~{\rm fm}^{-3}$.}} The critical mass, when the
branch of stable hybrid stars starts is $M_c^{\rm
quark}=1.214~M_\odot$.

As shown in Refs. \cite{Aguilera:2004ag,Blaschke:2005uj} quark
matter itself can exist in a mixed phase of 2SC and normal states.
The presence or absence of the 2SC - normal quark mixed phase
instead of only one of those phases is not so important for the
hybrid star cooling problem since the latter is governed by
processes involving either normal excitations or excitations with
the smallest gap.

\subsection{Stability of Hybrid star configurations}

  Not all compositions of hadronic and quark matter EoS's give
stable configu-rations for compact objects, which could be
demonstrated using the HHJ model EoS for hadronic and the two
flavor nonlocal chiral model EoS for quark matter.

 The mechanical structure of the spherically symmetric, static gravitational self-bound
configurations of dense matter can be calculated with the
well-known Tolman-Oppenheimer-Volkoff equations. These are the
conditions of hydrostatic equilibrium of self- gravitating matter,
see also \cite{Glendenning:wn},
\begin{equation}
\label{TOV} \frac{dP(r)}{dr}=
-\frac{[\varepsilon(r)+P(r)][m(r)+4\pi r^{3}P(r)]}{r[r-2m(r)]}~.
\end{equation}

Here $\varepsilon(r)$ is the energy density and $P(r)$ the
pressure at distance $r$ from the center of the star. The mass
enclosed in a sphere with radius $r$ is defined by
\begin{equation}
m(r)=4\pi \int_{0}^{r}\varepsilon(r')r'^{2}dr'~.
\end{equation}

These equations are solved for given central baryon number
densities, $n_B(r=0)$, thereby defining a sequence of star
configurations.

 The configurations are stable if they are on the
rising branch of the mass - central density relation.

The relation between pressure and energy density is given by the
choice of the corresponding EoS model, which generally can be a
function of temperature. The temperature is either constant for
isothermal configurations, see \cite{ST}, or has some profile,
when the cooling evolution is assumed to be a hydrodynamically
quasi-stationary process.

Hot quark stars have been discussed, e.g., in
\cite{Blaschke:1998hy,Blaschke:2003yn,Kettner:1994zs}.

For late cooling, when the temperatures are below $T < 1 ~{\rm
MeV}$ the effects of temperature on the distribution of matter are
negligible and the calculations for the star structure can be
performed for the $T = 0$ case.


Dots in Fig. \ref{fig3} indicate threshold densities for the DU
process. The possibility of charged pion condensation is
suppressed.  Otherwise for $n>n_c^{\rm DU}$ the isotopic
composition is changed in favor of increasing proton fraction and
a smaller critical density for the DU reaction. Deviations in the
$M(n)$ relation for HHJ and NLW EoS are minor, whereas the DU
thresholds are quite distinct.

\begin{figure}[ht]
\begin{center}
  \includegraphics[height=0.7\textwidth,angle=-90]{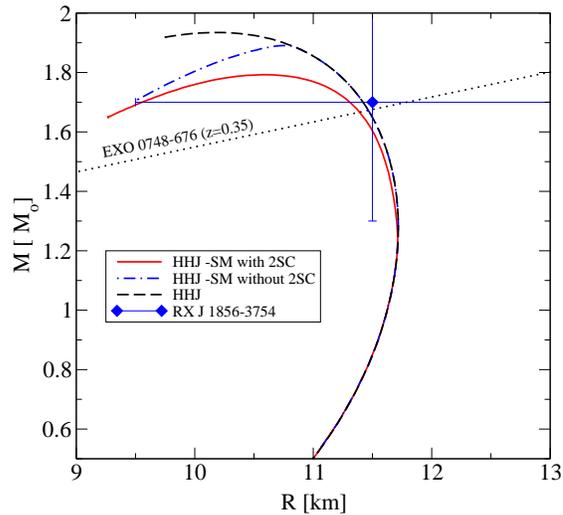}
   \caption{Mass -
radius relations for compact star configurations with different
EoS: purely hadronic star with HHJ EoS (dashed line), stable
hybrid stars with HHJ - Gaussian nonlocal chiral quark separable
model (SM) with 2SC phase (solid line) and with HHJ - SM, without
2SC phase (dash-dotted line). Data for two sources are also
indicated, see \cite{PLSP,CPM}. \label{fig:stab}}
\end{center}
\end{figure}

In Fig. \ref{fig:stab} we show the mass-radius relation for hybrid
stars with HHJ EoS vs. Gaussian nonlocal chiral quark separable
model (SM) EoS. Configurations with possible 2SC phase given by
the solid line, are stable, whereas  the hybrid star
configurations without 2SC large gap color superconductivity
(dash-dotted lines) are unstable. In case of  ``HHJ-SM with 2SC''
the maximum neutron star mass proves to be $1.793~M_{\odot}$. For
an illustration of the constraints on the mass-radius relation
which can be derived from compact star observations we show in
Fig. \ref{fig:stab} the compactness limits from the thermal
emission of the isolated neutron star RX J1856.5-3754 as given in
\cite{PLSP} and from the redshifted absorption lines in the X-ray
burst spectra of EXO 0748-676 given in \cite{CPM}. These are,
however, rather weak constraints.

\section{Heat transport, Neutrino production and diffusion}

\subsection{Boltzmann equation in curved space - time
}
  For study of cooling evolution, we will use the diffusion
approximation for neutrino transport, because it allows us easily
to assess how global characteristics of the neutrino emission such
as average energies, integrated fluxes, and time scales change in
response to various input parameters.  This approach is sufficient
to establish connections between the microphysical ingredients and
the duration of deleptonization and cooling time scales, which are
necessary to estimate possible effects on neutrino signals.
However, the detailed results of simulations for early cooling
will not be presented in this lecture. The reader is referred to
the works \cite{Pons:1998mm,Lindquist} and citations within.

We will assume that the neutron star has an interior where the
matter is spherically symmetric distributed. The evolution of the
star will be assumed to be quasi stationary, which means that for
each time step the velocity of the matter is zero.

 The Boltzmann equation (BE) for massless particles is
\begin{equation}
p^{\beta}{\left( {\frac {\partial f}{\partial x^{\beta}}} -
{{\Gamma}_{\beta\gamma}^{\alpha}}  p^{\gamma} {\frac {\partial
f}{\partial p^{\alpha}}} \right)} = {\left( \frac {df}{d\tau}
\right)}_{coll}\,
\end{equation}
where $f$ is the invariant neutrino distribution function,
$p^{\alpha}$  is the neutrino 4-momentum and
${\Gamma}^{\alpha}_{\beta \gamma}$  are the Christoffel symbols
for the metric
\begin{equation}
\label{metric}
d\tau^2=-e^{2\phi}dt^2+e^{2\Lambda}dr^2+r^2d\theta^2+r^2\sin^2\theta
\,d\Phi^2\,
\end{equation}
of static spherically symmetric space-time manifold.

For simplicity one can use the comoving basis and rewrite the BE
\begin{equation}
p^{b}{\left( {e^{\beta}_b} {\frac {\partial f}{\partial
x^{\beta}}} - {\Gamma}^a_{bc}  p^c {\frac {\partial f}{\partial
p^a}} \right)} = {\left( \frac {df}{d\tau} \right)}_{coll},
\end{equation}
where ${e^{\beta}_b}$ are basis vectors in a comoving frame to an
observer and in the static case they are diagonal. Indices $a,b,c$
are running from 0 to 3. The ${\Gamma}^a_{bc}$ are Ricci
rotational coefficients with the  non-zero components
\begin{eqnarray}
\Gamma^1_{00} &=&{\Gamma}^0_{01}=  e^{-\Lambda} \frac{\partial
\phi}{\partial r}, \\ \nonumber
 \Gamma^2_{21}&=& -{\Gamma}^1_{22} =  -{\Gamma}^1_{33} = {\Gamma}^3_{31}= {{e^{-\Lambda} }
 \over{r}},
\\ \nonumber
 {\Gamma}^2_{33} &=&-{\Gamma}^3_{32}= - \frac{\cot \theta}{r}.
\end{eqnarray}

The neutrino 4-momentum is
$$ p^a = \left(\omega, \omega{\mu}, \omega{(1-{\mu}^2)}^{1/2}\cos
{\Phi}, \omega{(1-{\mu}^2)}^{1/2}\sin {\Phi}\right) \,,
$$
where $\mu$  is the cosine of the angle between the neutrino
momentum and the radial direction, $\omega$  is the neutrino
energy in a comoving frame.
Using the previous definitions we will have
\begin{eqnarray}
\label{BE} \omega e^t_0 \frac{\partial f}{\partial t} &+& \omega
\mu e^r_1 \frac{\partial f}{\partial r} -\omega^2  {\mu}
\Gamma^1_{00} \frac{\partial f}{\partial \omega} - \omega {(1-
{\mu}^2)} {\left( \Gamma^1_{00} + \Gamma^1_{22} \right)}
\frac{\partial f}{\partial \mu} = {\left( \frac {df}{d \tau}
\right)}_{coll} \,.
\end{eqnarray}
\subsubsection{Equations in Linear response Approximation}

After applying the operator
$$
 \frac{1}{2} \int_{-1}^{+1} d{\mu}\, {\mu}^{i} \,, \quad
 i=0,1,2,\cdots, \;\;
$$
to equation (\ref{BE}) and defining the $i^{\rm th}$ moments
\cite{Thorne}
$$
 M_{i}=\frac{1}{2} \int_{-1}^{+1} d{\mu}\, {\mu}^{i} f, \;\;
 Q_i = \frac{1}{2} \int_{-1}^{+1}d{\mu}\, \mu^i {\left(\frac {df}{d
\tau}\right)}_{coll} \,. $$
we get for $i=0$
\begin{eqnarray}
\label{mom0} {\omega} \left({e^t_0} \frac{\partial M_0}{\partial
t} + {e^r_1} \frac{\partial M_1}{\partial r} \right) - {\omega}^2
{\left( \Gamma^1_{00} \frac{\partial M_1}{\partial {\omega}}
\right)} - 2 {\omega} (\Gamma^1_{00} + \Gamma^1_{22}) M_1  = Q_0,
\end{eqnarray}
and for $i=1$
\begin{eqnarray}
\label{mom1} {\omega} \left( {e^t_0} \frac{\partial M_1}{\partial
t} + {e^r_1} \frac{\partial M_2}{\partial r} \right) - {\omega}^2
{\left( \Gamma^1_{00} \frac{\partial M_2}{\partial {\omega}}
\right)} + {\omega} (\Gamma^1_{00} + \Gamma^1_{22})(M_0 - 3 M_2) =
Q_1 \,.
\end{eqnarray}

Let us introduce $N_{\nu}$, $F_{\nu}$, and $S_N$ as the number
density, number flux and the number source term, respectively,
\begin{equation}
\label{defs1} N_{\nu}= \int_0^{\infty} \frac{d\omega}{2 \pi^2} \,
M_0 \omega^2 \,, \quad F_{\nu}= \int_0^{\infty} \frac{d\omega}{2
\pi^2} \, M_1 \omega^2 \,, \quad S_N= \int_0^{\infty}
\frac{d\omega}{2 \pi^2} \, Q_0 \omega, \nonumber
\end{equation}
while $J_{\nu}$, $H_{\nu}$, $P_{\nu}$, and $S_E$ are the neutrino
energy density, energy flux, pressure, and the energy source
term
\begin{eqnarray}
\label{defs2}
J_{\nu}&=& \int_0^{\infty}
\frac{d\omega}{2 \pi^2} \, M_0 \omega^3 \,, \quad H_{\nu}=
\int_0^{\infty} \frac{d\omega}{2 \pi^2} \, M_1 \omega^3 \,, \nonumber\\
P_{\nu}&=& \int_0^{\infty} \frac{d\omega}{2 \pi^2} \, M_2 \omega^3
\,, \quad S_E= \int_0^{\infty} \frac{d\omega}{2 \pi^2} \, Q_0
\omega^2 \,.
\end{eqnarray}
 After integration over the neutrino energy and utilizing  the continuity equation
under assumption of a quasi-static evolution, one can recover the
well known neutrino transport equation \cite{Burrows}
\begin{equation}
 \label{number}
 \frac{\partial({N_{\nu}/n_B)}}{\partial t}
 +{\frac{\partial (e^{\phi} 4 \pi r^2 F_{\nu})}{\partial a}} =
e^\phi \frac{S_N}{n_B},
 \end{equation}

 \begin{equation}
 \label{energy}
\frac{\partial ({J_{\nu}/n_B})}{\partial t} + P_{\nu}
\frac{\partial ({1/n_B})}{\partial t} + e^{-\phi} {\frac{\partial
(e^{2 \phi} 4 \pi r^2 H_{\nu})}{\partial a}} = e^\phi
\frac{S_E}{n_B} \,.
\end{equation}

The distribution function in the diffusion approximation could be
represented in the following way
\begin{equation}
f(\omega,\mu)= f_0(\omega) +  \mu f_1(\omega) \,, \quad f_0 =
[1+e^{\left(\frac{\omega-\mu_\nu}{kT}\right)}]^{-1} \,,
\end{equation}
where $f_0(\omega)$  is the distribution function in equilibrium
($T=T_{mat}$, $\mu_{\nu}=\mu_{\nu}^{eq}$) with the neutrino energy
$\omega$ and the chemical potential $\mu_\nu$ respectively.
Hereafter the dependence of $f_0$ and $f_1$ on $\omega$ and non
explicit dependence on space-time coordinates will be assumed
without listing in arguments.

Thus the moments $M_i$ of the distribution function $f$ are
\begin{equation}
M_0= f_0 \,,\,\, M_1= \frac{1}{3}f_1 \,,\,\, M_2= \frac{1}{3}f_0
\,,\,\, \quad {\rm and} \quad M_3= \frac{1}{5}f_1 \,.
\end{equation}
Therefore the equation (\ref{mom1}) now reads
\begin{equation}
\label{mom1s} {e^{-\Lambda}} \left( \frac{\partial f_0}{\partial
r} - {\omega}{\frac{\partial \phi}{\partial r}} \frac{\partial
f_0}{\partial \omega} \right) = 3 \frac{Q_1}{\omega}\,.
\end{equation}
The collision term $Q_1$ can be represented as
\begin{equation}
\label{coll} {\left(\frac{df}{d\tau}\right)}_{coll} = \omega
\left( j_a(1-f) - \frac{f}{\lambda_a} + j_s(1-f) -
\frac{f}{\lambda_s} \right) \,.
\end{equation}
Here $j_a$  is the emissivity,  $\lambda_a$  is the absorptivity,
$j_s$  and  $\lambda_s$  are the scattering contributions.

Namely
\begin{equation}
\label{scat1} j_s = \frac{1}{{(2 \pi)}^3} \int_{0}^{\infty}
d\omega'\, \omega'^{2} \int_{-1}^1 d\mu' \int_0^{2\pi} d\Phi  \,
f(\omega',\mu') R^{in}_s(\omega,\omega', \cos\theta)\,,
\end{equation}
\begin{equation}
\label{scat2} \frac{1}{\lambda_s} = \frac{1}{{(2 \pi)}^3}
\int_{0}^{\infty} d\omega'\, \omega'^{2} \int_{-1}^1 d\mu'
\int_0^{2\pi} d\Phi  \, [1-f(\omega',\mu')]
R^{out}_s(\omega,\omega', \cos\theta) \,,
\end{equation}
where $\theta$ is the scattering angle.
The relation between emissivities and absorptivities ($R^{in}_s$
and $R^{out}_s$ are scattering kernels) is  given by
\begin{equation}
\frac{1}{\lambda_a(\omega)}=e^{\beta(\omega-\mu_{\nu}^{eq})}
j_a(\omega) \quad {\rm and} \quad
{R^{in}_s}=e^{\beta(\omega'-\omega)} R^{out}_s \,.
\end{equation}
Using the Legendre expansion for the moments one has
\begin{equation}
R^{out}_l = \int_{-1}^1 d\cos{\theta}\, P_l (\cos\theta)
R^{out}_s(\omega, \omega', \cos{\theta} ) \,.
\end{equation}

Performing the angular integrations of  Eq. (\ref{coll}) one can
define the relation between $Q_0$, $Q_1$ and $R^{out}_0$,
$R^{out}_1$.

After substitution of the expression for $Q_1$ into
Eq.(\ref{mom1s}) the relation between $f_0$ and $f_1$ reads

\begin{eqnarray}
\label{f1} f_1 = - D(\omega) \left[ \frac{\partial f_0}{\partial
r} - \omega  \frac{\partial \phi}{\partial r} {\frac{\partial
f_0}{\partial \omega}} \right]e^{-\Lambda} \,, D(\omega) = {\left(
j_a +\frac{1}{\lambda_a}+\kappa^s_1 \right)}^{-1} \,,
\end{eqnarray}
where $ D(\omega)$ is the diffusion coefficient. Using the relation
between $\frac{\partial f_0}{\partial r}$ and $\frac{\partial
f_0}{\partial \omega}$
and the notation $ \eta=\mu_{\nu}/T $ to be the neutrino
degeneracy parameter, we obtain
\begin{equation} f_1 = - D(\omega) e^{-\Lambda} \left[ T
\frac{\partial \eta}{\partial r} + \frac{\omega}{T e^{\phi}}
\frac{\partial (T e^{\phi})}{\partial r} \right] \left(- \,
\frac{\partial f_0}{\partial \omega} \right)\,.
\end{equation}
Now the energy-integrated lepton and energy fluxes are
\begin{eqnarray}
 F_{\nu}&=&  - \, \frac{e^{-\Lambda} e^{-\phi}T^2}{6 \pi^2}
\left[ D_3 \frac{\partial (T e^{\phi})}{\partial r} + (T e^{\phi})
D_2 \frac{\partial \eta}{\partial r}  \right] \nonumber \\
H_{\nu}&=&- \, \frac{e^{-\Lambda} e^{-\phi}T^3}{6 \pi^2} \left[
D_4 \frac{\partial (T e^{\phi})}{\partial r} + (T e^{\phi}) D_3
\frac{\partial \eta}{\partial r}  \right] \,. \label{fluxes}
\end{eqnarray}
Here the coefficients $D_2$, $D_3$, and $D_4$ are defined through
the integrals over $\omega$ as
$$
D_n = T^{-(n+1)}\,\int_0^\infty d\omega~\omega^n
D(\omega)f_0(\omega)(1-f_0(\omega))~. \label{d2d3}
$$
The explicit expression for the absorption mean free path is
\begin{eqnarray}
\frac{1}{\lambda_a} =  \frac{G_F^2}{\pi^2} \int_{0}^{\infty} dE_e
\, E_e^2 \, \left[1-f_{eq}(E_e)\right] \int_{-1}^{+1}
d\cos{\theta} \, \frac{(\cos\theta-1)}{1-e^{-z}} \left[ A R_1 +
R_2 + B R_3 \right]
\end{eqnarray}
\begin{equation}
R^{out}_s = {4 G_F^2} \frac {(\cos\theta-1)}{1-e^{-z}} \left[ A
R_1 + R_2 + B R_3 \right]
\end{equation}
The response functions written in terms of the polarization functions are
\begin{eqnarray}
R_1  &=&  ({\cal V}^2+{\cal A}^2) ~[{\rm Im}~ \Pi^R_L(q_0,q)+{\rm
Im}~ \Pi^R_T(q_0,q)]\\
 R_2   &=&   ({\cal V}^2 + {\cal A}^2)~{\rm
Im}~ \Pi^R_T(q_0,q) - {\cal A}^2~{\rm Im}~ \Pi^R_A(q_0,q)
\\
\;\;\;\;R_3   &=&   2{\cal V}{\cal A} ~{\rm Im}~ \Pi^R_{VA}(q_0,q)
\,,
\end{eqnarray}
where the coupling constants for the absorption are ${\cal V}=g_V
\cos (\theta_c) $ and ${\cal A}=g_A\cos (\theta_c)$ and for the
scattering ${\cal V}=g_V/2$ and ${\cal A}=g_A/2$. $\theta_c $ is
the Cabibbo angle, $g_V$ and $g_A$ are the vector and axial-vector
couplings.

Introducing the $s$ entropy per baryon and $Y_L = (N_\nu + N_e +
N_\mu)/n_B$ lepton number fraction and with help of conservation
laws for energy and baryon number combined with Eqs.
(\ref{number}) and (\ref{energy}) we finally will obtain the
energy transport equation
%
\begin{equation}
\label{enf}
 { T} e^{\phi}\frac{\partial{{s}}}{\partial{t}}+
{ \mu_{\nu}} e^{\phi}\frac{\partial{{
Y_L}}}{\partial{t}}+\frac{\partial{(e^{2\phi}
    4\pi r^2 { H_{\nu}})}}{\partial{a}}=0
\end{equation}
and the equation for lepton diffusion
\begin{eqnarray}
\frac{\partial{{Y_L}}}{\partial{t}}+\frac{\partial{(e^{\phi}
    4\pi r^2 {F_{\nu}})}}{\partial{a}}=0.
\end{eqnarray}
The energy $H_\nu$ and lepton $F_\nu$ fluxes are defined in
Eq.(\ref{fluxes}).

These coupled equations for the unknowns $T$ and $\mu_\nu$ with
the Eq.(\ref{TOV}) for the star structure completely describe the
early cooling evolution of the compact objects in quasi-stationary
regime. The solutions and more detailed information one can find
in \cite{Pons:1998mm}. In the next section we will use the energy
transport equation for consideration of late cooling evolution
when the neutrinos are already untrapped, i.e. the mean free path
$\lambda_a$ is larger than the star radius $R$ and $\mu_\nu = 0$.


\subsection{Late cooling evolution}

The cooling scenario is described with a set of cooling
regulators. In the regime without neutrino diffusion in the
righthand side of Eq. (\ref{energy}) and therefore also in Eq.
(\ref{enf}) for energy transport we introduce the energy loss term
$\epsilon_\nu$ due to neutrino emission additional to $S_E$ and
instead of the $H_\nu$ we use the total energy flux function
$l(r,t)$.  The temperature profile $T(r,t)$ is the one time
dependent unknown dynamical quantity and during the cooling
process due to an inhomogeneous distribution of the matter inside
the star and the finite heat conductivity it can differ from the
isothermal one.  The heat conductivity $\kappa$ (the corresponding
term to $D_4$ in Eq. (\ref{enf})) determines the relation between
$l(r,t)$ and gradient of $T(r,t)$. In isothermal age the
temperature on the inner crust boundary ($T_{m}$) and the central
temperature ($T_c$) are connected by the relation $T_{m} = T_{c}
\exp [\phi(0) - \phi(R)]$. Here $\phi(0)-\phi(R)$ is the
difference of the gravitational potentials in the center and at
the surface of the star, respectively.

The flux of energy $l(r)$ per unit time through a spherical slice
at the distance $r$ from the center, is proportional to the
gradient of the temperature on both sides of this slice,
\begin{equation}\label{lr}
l(r) = - 4 \pi r^2 \kappa (r) \frac{\partial (T{\rm
e}^\phi)}{\partial r} {\rm e}^{-\phi} \sqrt{1-\frac{2 M}{r}}~,
\end{equation}
where the factor ${\rm e}^{-\phi} \sqrt{1-\frac{2 M}{r}}$
corresponds to the relativistic correction of the time scale and
the unit of thickness. The equations for energy balance and
thermal energy transport are \cite{W99}
\begin{eqnarray}\label{lbal}
\frac{\partial }{\partial A}\left( l {\rm e}^{2 \phi}\right)&=& -
\frac{1}{n}\left( \epsilon_\nu {\rm e}^{2\phi}
+ c_V \frac{\partial }{\partial t} (T {\rm e}^\phi)\right) ~,\\
\frac{\partial }{\partial A}\left(T {\rm e}^{\phi}\right)&=& -
\frac{1}{\kappa} \frac{l {\rm e}^{\phi}}{16 \pi ^2 r^4 n} ~,
\label{Tbal}
\end{eqnarray}
where $n=n(r)$ is the baryon number density, $A=A(r)$ is the total
baryon number within a sphere of radius $r$. One has
\begin{equation}
\frac{\partial r}{\partial A}=\frac{1}{4 \pi r^2 n}
\sqrt{1-\frac{2 M}{r}}~.
\end{equation}
The total neutrino emissivity $\epsilon_\nu$ and the total
specific heat $c_V$ are given as the sum of the corresponding
partial contributions defined in the next subsections. The density
profiles $n_i(r)$ of the constituents $i$ of the matter are under
the conditions of the actual temperature profile $T(r,t)$. The
accumulated mass $M=M(r)$ and the gravitational potential
$\phi=\phi(r)$ can be determined by
\begin{eqnarray}\label{potential}
\frac{\partial M}{\partial A}&=&\frac{\varepsilon}{n}
\sqrt{1-\frac{2 M}{r}}~,
\\
\frac{\partial \phi}{\partial A}&=&\frac{4 \pi r^3 p + M}{4 \pi
r^2 n} \frac{1}{\sqrt{1-\frac{2 M}{r}}}~,
\end{eqnarray}
where the energy density profile $\varepsilon=\varepsilon(r)$ and
the pressure profile $p=p(r)$ is defined by the condition of
hydrodynamical equilibrium (see Eq. \ref{TOV})
\begin{equation}\label{tov}
\frac{\partial p}{\partial A}= - (p + \varepsilon) \frac{\partial
\phi}{\partial A}~.
\end{equation}
The boundary conditions for the solution of (\ref{lbal}) and
(\ref{Tbal}) read $l(r=0)=l(A=0)=0$ and
$T(A(r_m)=A,t)=T(r_m=R,t)=T_m(t)$, respectively.

In our examples we choose the initial temperature to be 1 MeV.
This is a typical value for the temperature $T_{\rm opacity}$ at
which the star becomes transparent for neutrinos. Simplifying we
disregard the neutrino influence on transport. These effects
dominate for $t < 1 \div 100$ min, when the star cools down to $T
\leq T_{\rm opacity}$ and become unimportant for later times.

\subsubsection{ Neutrino Processes in dense matter}

We compute the NS thermal evolution adopting our fully general
relativistic evolutionary code. This code was originally
constructed for the description of  hybrid stars by
\cite{Blaschke:2000dy}. The main cooling regulators are the
thermal conductivity, the heat capacity and the emissivity.

\subsubsection*{Emissivity}

The luminosities are calculated using the corresponding matrix
element of the neutrino production process
\begin{equation}
 L_\nu=(2\pi)^4\int
\frac{d^3p_n}{(2\pi)^3 2 E_n}\dots \int \frac{d^3p_\nu}{(2\pi)^3 2
E_\nu}~\delta^3(\vec{p}_i)\delta(E_i) ~\big| M_{fi}\big|^2 ~
f_n(1-f_p)(1-f_e).
\end{equation}

The most effective process of neutrino production is the direct
Urca (DU) one, $n\longrightarrow p + e^- + \bar\nu_e$, which is
allowed when the momentum conservation holds
~$\vec{p}_{F,n}=\vec{p}_{F,p}+\vec{p}_{F,e}$ ~ $\Leftrightarrow$ ~
$|\vec{p}_{F,n}|\le |\vec{p}_{F,p}|+|\vec{p}_{F,e}|$. Since DU
process is so intensive that as soon as it takes place the star
cools too fast for better correspondence between the cooling
simulations and the observational data, it is likely to assumed DU
to be forbidden in the interior of the star.

\begin{figure}[ht]
\begin{center}
\includegraphics[height=0.7\textwidth,angle=-90]{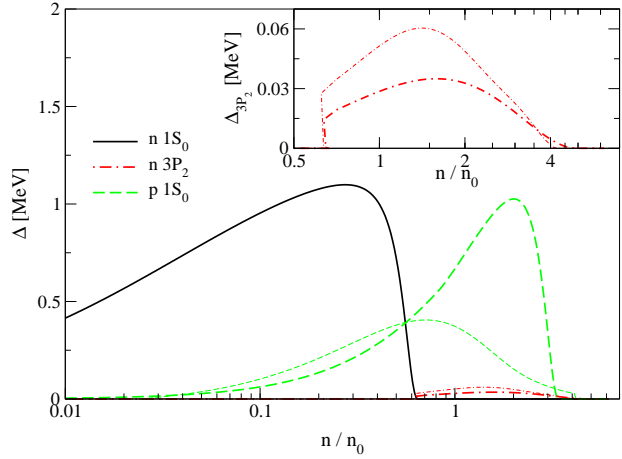}
\caption{ Neutron and proton pairing gaps according to model I
(thick solid, dashed and dotted lines) and according to model II
(thin lines), see text. The $1S_0$ neutron gap is the same in both
models, taken from  \cite{AWP}. \label{fig-gaps}}
\end{center}
\end{figure}

  The other processes, i.e. the modified Urca (MU), pair breaking and
formation (PBF) and bremsstrahlung are the main ones governing the
cooling in hadronic matter. The modified Urca (MU) $nn\rightarrow
npe\bar{\nu}$, $np\rightarrow ppe\bar{\nu}$, has to be medium
modified Urca (MMU) process due to the softening of in-medium pion
propagator. The final emissivity is given by \cite{FM79,YLS99}
\begin{eqnarray}
\label{eq:1} \epsilon_{\nu}^{\rm nMU}&=&8.6\times
10^{21}m^{*4}_{\rm nMU}
(Y_e u)^{1/3}\zeta_{\rm nMU}T_9^8 {\rm erg cm^{-3}s^{-1}},\\
\epsilon_{\nu}^{\rm pMU}&=&8.5\times 10^{21}m^{*4}_{\rm pMU} (Y_e
u)^{1/3}\zeta_{\rm pMU}T_9^8 {\rm erg cm^{-3}s^{-1}}.
\end{eqnarray}
Here $m_i^* =\sqrt{m_{{\rm{rel}},i}^{*2}+p_{{\rm{F}},i}^2}$ is the
non-relativistic quasiparticle effective mass related to the
in-medium one-particle energies from a given relativistic mean
field model for $i=n,p$. We have introduced the abbreviations
$m^{*4}_{\rm nMU}=({m_n^*}/{m_n})^3({m_p^*}/{m_p})$ and
$m^{*4}_{pMU}=({m_p^*}/{m_p})^3({m_n^*}/{m_n})$. The suppression
factors are $\zeta_{\rm nMU}=\zeta_n \zeta_p \simeq
{\exp}\{-[\Delta_n(T)+\Delta_p(T)]/T\}$, $\zeta_{\rm pMU}\simeq
\zeta_p^2$, and should be replaced by unity for $T>T_{{\rm
crit},i}$, when for given species $i$ the corresponding gap
vanishes. For neutron and proton $S$-wave pairing is
$\Delta_{i}(0)=1.76~T_{{\rm crit},i}$ and for the $P$-wave pairing
of neutrons $\Delta_n(0) =1.19~ T_{{\rm crit},n}$ (see Fig.
\ref{fig-gaps}). The gap as a function of temperature is given by
the interpolation formula $\Delta_N(T)=\Delta(0) \sqrt{1-T/T_{{\rm
crit},N}}$.

To be conservative we have used in (\ref{eq:1}) the free one-pion
exchange estimate of the $NN$ interaction amplitude. Restricting
ourselves to a qualitative analysis we use here simplified
exponential suppression factors $\zeta_i$. In a more detailed
analysis these $\zeta_i$-factors have prefactors with rather
strong temperature dependences \cite{YLS99}. At temperatures
$T\sim T_c$ their inclusion only slightly affects the resulting
cooling curves. For $T\ll T_c$ the MU process gives in any case a
negligible contribution to the total emissivity and thereby
corresponding modifications can again be omitted. Also for the
sake of simplicity the general possibility of a $^3$P$_2(|m_J|=2)$
pairing which may result in a power-law behaviour of the specific
heat and the emissivity of the MU process
\cite{Sedrakian:2004qd,YLS99,VS87} is disregarded since mechanisms
of this type of pairing are up to now not elaborated. Even more
essential modifications of the MU rates may presumably come from
in-medium effects which could result in extra prefactors of $10^2
\div 10^3$ already at $T \sim T_c$.

In order to estimate the role of the in-medium effects in the $NN$
interaction for the HNS cooling we have also performed
calculations for the so-called medium modified Urca (MMU) process
\cite{VS86,MSTV90} by multiplying the rates (\ref{eq:1}) by the
appropriate prefactor
\begin{equation}\label{VS-FM}
\epsilon_{\nu}^{\rm MMU}/\epsilon_{\nu}^{\rm MU}\simeq 10^{3}
\left[ \Gamma^{6}(g^{\prime})/\widetilde{\omega}^{8}(k\simeq p_F
)\right] u^{10/3},
\end{equation}
where the value $\Gamma (g^{\prime}) \simeq 1/[1+1.4~u^{1/3}]$ is
due to the dressing of $\pi NN$ vertices and
$\widetilde{\omega}\leq m_\pi$ is the effective pion gap which we
took as function of density from Fig. 2 of \cite{SVSWW97}.

For $T<T_{\rm crit}$ the most important contribution comes from
the neutron \cite{VS87,YLS99} and the proton \cite{VS87} pair
breaking and formation processes. We take their emissivities from
Ref. \cite{SVSWW97}, which is applicable for both cases, $S$- and
$P$-wave nucleon pairing
\begin{eqnarray}
\label{eq:2} \epsilon_{\nu}^{\rm nPBF}&=&6.6\times 10^{28}
({m_n^*}/{m_n})
({\Delta_n(T)}/{\rm MeV})^7~u^{1/3}\nonumber \\
&&\times \xi~I({\Delta_n(T)}/{T})~{\rm erg~ cm^{-3}s^{-1}},\\
\epsilon_{\nu}^{\rm pPBF}&=&0.8\times 10^{28} ({m_p^*}/{m_p})
({\Delta_p(T)}/{\rm MeV})^7~u^{2/3}\nonumber \\
&&\times ~I({\Delta_p(T)}/{T})~{\rm erg~cm^{-3}s^{-1}},
\end{eqnarray}
where
\begin{equation}
I({\Delta_i(T)}/{T})\simeq 0.89~\sqrt{T/\Delta_i(T)} \exp[-2
\Delta_i(T)/T]~, \label{eq:3}
\end{equation}
$\xi \simeq 0.5$ for $^1S_0$ pairing and $\xi \simeq 1$ for
$^3P_2$ pairing.

A significant contribution of the proton channel is due to the
$NN$ correlation effects, which are taken into account in \cite{VS87}.

The phonon contribution to the emissivity of the $3P_2$ superfluid
phase is negligible. The main emissivity regulators are the MMU
Eq. (\ref{VS-FM}) and the neutron (nPBF) and proton (pPBF) pair
breaking and formation  processes Eq. (\ref{eq:2}), see above for
a rough estimation. Finally, we include the effect of a pion
condensate (PU process), where the PU emissivity is about 1-2
orders of magnitude smaller than the DU one.

All emissivities are corrected by correlation effects. We adopt
the same set of partial emissivities as in the work of
\cite{SVSWW97}.

In quark matter the process $d\longrightarrow u + e^- +
\bar\nu_e$ is always allowed unless it is suppressed by huge
energy gaps due to quark pairing.

\begin{eqnarray}\label{neutr-DU}
\epsilon^{\rm QDU}_\nu \simeq 9.4 \times 10^{26}\alpha_s u
Y_e^{1/3}\zeta_{\rm QDU}~ T_{9}^6~ {\rm erg~cm^{-3}~s^{-1}},
\end{eqnarray}
where at a compression $u=n/n_0\simeq 2$ the strong coupling
constant is $\alpha_s \approx 1$ and decreases logarithmically at
still higher densities. The nuclear saturation density is
$n_0=0.17~ {\rm fm}^{-3}$, $Y_e =n_e /n$ is the electron fraction,
and $T_9$ is the temperature in units of $10^9$ K. If, for a
somewhat higher density, the electron fraction was too small
($Y_e<Y_{ec}\simeq 10^{-8}$), then all the QDU processes  would be
completely switched off \cite{DSW83} and the neutrino emission
would be governed by two-quark reactions like the quark modified
Urca (QMU) and the quark bremsstrahlung (QB) processes
$dq\rightarrow uqe\bar{\nu}$ and $q_1 q_2 \rightarrow q_1 q_2
\nu\bar{\nu}$, respectively. The emissivities of the QMU and QB
processes have been estimated as \cite{I82}
\begin{eqnarray}\label{neutr-B}
\epsilon^{\rm QMU}_\nu\sim \epsilon^{\rm QB}_\nu  \simeq 9.0\times
10^{19}\zeta_{\rm QMU}~T_{9}^8~ {\rm erg~cm^{-3}~s^{-1}}.
\end{eqnarray}
Due to the pairing, the emissivities of QDU processes are
suppressed by a factor $\zeta_{\rm QDU} \sim \mbox{exp}(-\Delta_q
/T)$ and the emissivities of QMU and QB processes are suppressed
by a factor $\zeta_{\rm QMU} \sim \mbox{exp}(-2\Delta_q /T)$ for
$T<T_{{\rm crit},q}\simeq 0.4~\Delta_q$ whereas for $T>T_{{\rm
crit},q}$ these factors are equal to unity. The modification of
$T_{{\rm crit},q}(\Delta_q )$ relative to the standard BCS formula
is due to the formation of correlations as, e.g., instanton-
anti-instanton molecules \cite{r+99}. For the temperature
dependence of the gap below $T_{{\rm crit},q}$ we use the
interpolation formula $\Delta(T)=\Delta(0) \sqrt{1-T/T_{{\rm
crit},q}}$, with $\Delta(0)$ being the gap at zero temperature.

The contribution of the reaction $ee\rightarrow ee\nu\bar{\nu}$ is
very small \cite{KH99}
\begin{equation}
\epsilon^{ee}_\nu = 2.8\times 10^{12}\, Y_e^{1/3} u^{1/3} T_9^8~
{\rm erg~cm^{-3}~s^{-1}}, \label{eq:4}
\end{equation}
but can become important, when quark processes are blocked out
for large values of $\Delta_q/T$ in superconducting quark matter.

\subsubsection*{Thermal conductivity}

The heat conductivity of the matter is the sum of the partial
contributions \cite{flowers,BH99}
\begin{equation}
 \label{conductivity2}
\kappa = \sum_{i}\kappa_{i},\,\,\, \frac{1}{\kappa_i} =
\sum_{j}\frac{1}{\kappa_{ij}}~,
\end{equation}
where $i,j$ denote the components (particle species).

The contribution of neutrons and protons is
\begin{eqnarray}
\label{kapn}
 \kappa_{nn}&=&8.3\times 10^{22}
\left(\frac{m_n}{m_n^*}\right)^4
 \frac{z_n^3~\zeta_n}{S_{kn}~T_9}~
{\rm erg~ s^{-1}cm^{-1}K^{-1}}, \\[8pt]
S_{kn}&=&0.38~z_n^{-7/2} + 3.7~z_n^{2/5}~,
\\
\kappa_{np}&=&8.9\times 10^{16} \left(\frac{m_n}{m_n^*}\right)^2
\frac{z_n^2\zeta_{p}T_9}{z_p^{3}S_{kp}}~{\rm erg~s^{-1}cm^{-1}K^{-1}},\\
S_{kp}&=&1.83~z_n^{-2} + 1.43~z_n^2(0.4+z_n^8)^{-1}~.
\label{kapp}
\end{eqnarray}

Here we have introduced the appropriate suppression factors
$\zeta_i$ which act in the presence of gaps for superfluid
hadronic matter (see Fig. \ref{fig-gaps}) and we have used the
abbreviation $z_i=(n_i/(4 n_0))^{1/3}$. The heat conductivity of
electrons is given by Eq. (\ref{ee}). The total contribution
related to electrons is then
\begin{eqnarray}\label{kapel}
1/\kappa_e = 1/\kappa_{ee}+1/\kappa_{ep}.
\end{eqnarray}
Similar expressions we have for neutrons and protons using Eqs.
(\ref{kapn})-(\ref{kapp}).

The total thermal conductivity is the straight sum of the partial
contributions $\kappa_{tot} = \kappa_{e}+\kappa_{n}+...$ . Other
contributions to this sum are smaller than those presented
explicitly ($\kappa_e$ and $\kappa_n$).

For quark matter $\kappa$ is the sum of the partial conductivities
of the electron, quark and gluon components \cite{BH99,HJ89}
\begin{equation}
 \label{total}
\kappa = \kappa_{e} + \kappa_{q}+\kappa_{g},
\end{equation}
where $\kappa_{e}\simeq \kappa_{ee}$ is determined by
electron-electron scattering processes since in superconducting
quark matter the partial contribution $1/\kappa_{eq}$ (as well as
$1/\kappa_{gq}$ ) is additionally suppressed by a $\zeta_{\rm
QDU}$ factor, as for the scattering on impurities in metallic
superconductors. For $\kappa_{ee}$ we have
\begin{eqnarray}
\label{ee}
\kappa_{ee}&=&5.5\times 10^{23}u~{Y_e}~{T_9}^{-1}~ {\rm
erg~s^{-1}cm^{-1}K^{-1}},
\end{eqnarray}
and
\begin{eqnarray}
\kappa_q \simeq  \kappa_{qq} \simeq& 1.1\times
10^{23}\sqrt{\frac{4\pi}{\alpha_s}}~u~\zeta_{\rm QDU}{T_9}^{-1}
{\rm erg~s^{-1}cm^{-1}K^{-1}},
\end{eqnarray}
where we take into account the suppression factor. We estimate the
contribution of massless gluons as
\begin{eqnarray}
\kappa_{g}\simeq \kappa_{gg} \simeq 6.0 \times
10^{17}~T_9^{2}~{\rm erg~s^{-1}cm^{-1}K^{-1}}.
\end{eqnarray}

\subsubsection*{Heat capacity}

The heat capacity contains nucleon, electron, photon, phonon, and
other contributions. The main in-medium modification of the
nucleon heat capacity is due to the density dependence of the
effective nucleon mass. We use the same expressions as
\cite{SVSWW97}. The main regulators are the nucleon and the
electron contributions. For the nucleons ($i=n,p$), the specific
heat is \cite{FM79}
\begin{equation}
c_i \sim 1.6 \times 10^{20}({m_i^*}/{m_i})~(n_i/n_0)^{1/3}
\zeta_{ii}~T_9~ {\rm erg~cm^{-3}K^{-1}}~,
\end{equation}
and for the electrons it is
\begin{equation}\label{e}
c_e \sim 6\times 10^{19} \,(n_e/n_0)^{2/3}~T_9~ {\rm
erg~cm^{-3}~K^{-1}}~ .
\end{equation}
Near the phase transition point the heat capacity acquires a
fluctuation contribution. For the first order pion condensation
phase transition this additional contribution contains no
singularity, unlike the second order
phase transition, see \cite{MSTV90}. Finally, the nucleon
contribution to the heat capacity may increase up to several times
in the vicinity of the pion condensation point. The effect of this
correction on global cooling properties is rather unimportant.

The symmetry of the $3P_2$ superfluid phase allows a Goldstone
boson (phonon) contribution of
\begin{eqnarray}
c_{G}\simeq 6\cdot 10^{14}T_9^3 \,\,\, {\rm erg}\,{\rm
cm}^{-3}~{\rm K^{-1}},
\end{eqnarray}
for $T<T_{cn}(3P_2)$, $n>n_{cn}(3P_2)$. We include this
contribution in our study too, although its effect on the cooling is
rather minor.

For the quark specific heat we use the expression \cite{I82}
\begin{eqnarray}\label{heat}
c_{q}\simeq 10^{21}u^{2/3}\zeta_{\rm S}~T_9~{\rm
erg~cm^{-3}~K^{-1}},
\end{eqnarray}
where $\zeta_{\rm S}\simeq 3.1~(T_{{\rm
crit},q}/T)^{5/2}~\mbox{exp} (-\Delta_q /T)$. Besides, one should
add the gluon-photon contribution \cite{Blaschke:1999qx}
\begin{eqnarray}\label{gluon}
c_{g-\gamma}= 3.0\times 10^{13}~N_{g-\gamma}~T_9^3~{\rm
erg~cm^{-3}~K^{-1}},
\end{eqnarray}
where $N_{g-\gamma}$ is the number of available  massless
gluon-photon states (which are present even in the color
superconducting phase), as well as the electron contribution,
\begin{equation}\label{e}
c_e= 5.7\times 10^{19} \,Y_e^{2/3} u^{2/3}~T_9~{\rm
erg~cm^{-3}~K^{-1}}.
\end{equation}

\section{Results of simulations for Cooling evolution}

We compute the neutron star thermal evolution adopting our fully
general relativistic evolutionary code. The code originally
constructed for the description of hybrid stars by
\cite{Blaschke:2000dy} has been developed and updated according to
the modern knowledge of inputs in \cite{Grigorian:2004jq}.



The density $n\sim 0.5\div 0.7 ~n_0$ is the boundary of the
neutron star interior and the inner crust. The latter is
constructed of a pasta phase discussed by \cite{RPW83}, see also
recent works of \cite{Maruyama:2004vy,Tatsumi:2005qn}.

Further on we need a relation between the crust and the surface
temperature for neutron star. A sharp change of the temperature
occurs in the envelope. This $T_{\rm s} - T_{\rm in}$ relation has
been calculated in several works, see \cite{GS80,Yakovlev:2003ed},
depending on the assumed value of the magnetic field at the
surface and some uncertainties in our knowledge of the structure
of the envelope. For applications we use three different
approximations: the simplified ``Tsuruta law'' $ T_{\rm s}^{\rm
Tsur}=(10 ~T_{\rm in})^{2/3}$ used in many old cooling
calculations, models used by Ref. \cite{Yakovlev:2003ed}, and our
fit formula interpolating two extreme curves describing the
borders of region of available $T_{\rm s} - T_{\rm in}$ relation,
taken from \cite{Yakovlev:2003ed} (see Fig. 4 of Ref.
\cite{Blaschke:2004vq}).

\subsection{Cooling Evolution of Hadronic Stars}

Here we will shortly summarize the results on hadronic cooling.

In framework of a ''minimal cooling'' scenario, the pair breaking and
formation (PBF) processes may allow to cover an ''intermediate
cooling'' group of data (even if one artificially suppressed
medium effects)\cite{SVSWW97}. These processes are very efficient
for large pairing gaps and temperatures being not much less than
the value of the gap.

Gaps, which we have adopted in the framework of the ''nuclear medium
cooling'' scenario, see \cite{Blaschke:2004vq}, are presented in
Fig. \ref{fig-gaps}. Thick dashed lines show proton gaps which
were used in the work of \cite{Yakovlev:2003qy} performed in the
framework of the ``standard plus exotics'' scenario. We will call
the choice of the ``3nt'' model from \cite{Yakovlev:2003qy} the
model I. Thin lines show $1S_0$ proton and $3P_2$ neutron gaps
from \cite{Takatsuka:2004zq}, for the model AV18 by \cite{WSS95}
(we call it the model II). Recently \cite{Schwenk:2003bc} has
argued for a strong suppression of the $3P_2$ neutron gaps, down
to values  $ \sim 10~$ keV, as the consequence of the
medium-induced spin-orbit interaction.

These findings motivated \cite{Blaschke:2004vq} to suppress values
of $3P_2$ gaps shown in Fig. \ref{fig-gaps} by an extra factor
$f(3P_2 ,n)=0.1$. Further possible suppression of the $3P_2$ gap
is almost not reflected on the behavior of the cooling curves.

Contrary to expectations of \cite{Schwenk:2003bc} a more recent
work of \cite{Khodel:2004nt} argued that the
$3P_2$ neutron pairing gap should be dramatically enhanced, as the
consequence of  the strong softening of the pion propagator.
According to their estimate, the
$3P_2$ neutron pairing gap is  as large as $1\div 10$~MeV in a
broad region of densities, see Fig. 1 of their work. Thus results
of calculations of \cite{Schwenk:2003bc} and \cite{Khodel:2004nt},
which both had the same aim to include medium effects in the
evaluation of the $3P_2$ neutron gaps, are in a deep discrepancy
with each other.



\begin{figure}[htb]
\begin{center}
\includegraphics[height=0.65\textwidth,angle=-90]{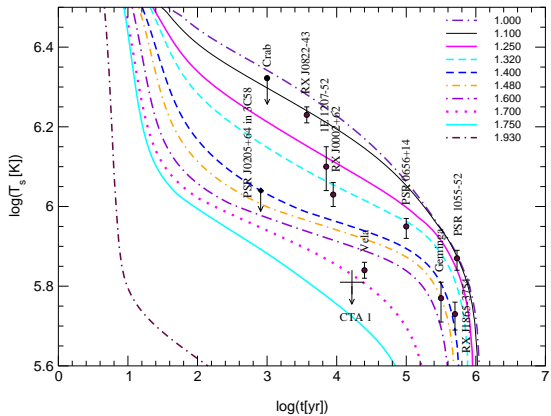}
\caption{ Cooling curves for the gaps from Fig. \ref{fig-gaps}
for model II (the same as Fig. 21 of \cite{Blaschke:2004vq}). The
original $3P_2$ neutron pairing gap is  additionally suppressed by
a factor $f(3P_2 ,n)=0.1$.
The $T_{\rm s} - T_{\rm in}$ relation is given by ``our fit''
curve of Fig. 4 in \cite{Blaschke:2004vq}.
For more details see \cite{Blaschke:2004vq}. \label{fig21BGV}}
\end{center}
\end{figure}
\begin{itemize}

\item
Including superfluid gaps we see, in agreement with recent
microscopic findings of \cite{Schwenk:2003bc},  that the $3P_2$
neutron gap should be as small as $ 10$~keV or less. So the
``nuclear medium cooling'' scenario of \cite{Blaschke:2004vq}
supports results of \cite{Schwenk:2003bc} and fails to
appropriately fit the neutron star cooling data when a strong
enhancement of the $3P_2$ neutron gaps is assumed as shown by
\cite{Grigorian:2005fi}.


\item
 Medium effects associated with the pion softening are called for
by the data. As the result of the pion softening the pion
condensation may occur for $n\geq n_c^{\rm PU}$ ($n\geq 3n_0$ in
our model). Its appearance at such rather high densities does not
contradict to the cooling data (see Fig. \ref{fig21BGV}), but also
the data are well described using the  pion softening but without
assumption on the pion condensation. The similar argumentation
holds also for DU threshold density. That puts restrictions on the
density dependence of the symmetry energy. Both  statements might
be important in the discussion of the heavy ion collision
experiments.

\item
We demonstrated a regular mass dependence for neutron stars with
masses $M> 1~M_{\odot}$, where less massive neutron stars cool
down slower and  more massive neutron stars cool faster. This
feature is more general and concerning also to hybrid stars
\cite{Grigorian:2005ds}.
\end{itemize}

\subsection{Cooling Evolution of Hybrid Stars with 2SC Quark Matter Core}


For the calculation of the cooling of the quark core in the hybrid
star we use the model \cite{Grigorian:2004jq}. We incorporate the
most efficient processes: the quark direct Urca (QDU) processes on
unpaired quarks, the quark modified Urca (QMU), the quark
bremsstrahlung (QB), the electron bremsstrahlung (EB), and the
massive gluon-photon decay (see \cite{Blaschke:1999qx}). Following
\cite{Jaikumar:2001hq} we include the emissivity of the quark pair
formation and breaking (QPFB) processes, too. The specific heat
incorporates the quark contribution, the electron contribution and
the massless and massive gluon-photon contributions. The heat
conductivity contains quark, electron and gluon terms.

The calculations are based on the hadronic cooling scenario
presented in Fig. \ref{fig21BGV} and we add the contribution of
the quark core. For the Gaussian form-factor the quark core occurs
already for $M>1.214~M_\odot$ according to the model
\cite{Blaschke:2003yn}, see Fig. \ref{fig:stab}. Most of the
relevant neutron star configurations (see Fig. \ref{fig21BGV}) are
then affected by the presence of the quark core.

\begin{figure}
\begin{center}
\includegraphics[height=0.65\textwidth,angle=-90]{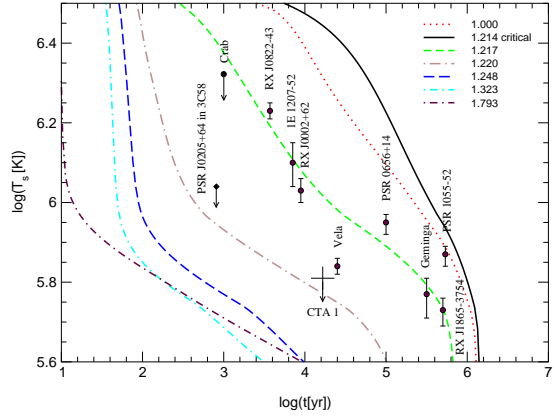}
\caption{Model I. Cooling curves for hybrid star configurations
with Gaussian quark matter core in the 2SC phase. The labels
correspond to the gravitational masses of the configurations in
units of the solar mass.} \label{fig:cool-2sc}
\end{center}
\end{figure}

First we check the possibility of the 2SC+ normal quark phases,
see Fig. \ref{fig:cool-2sc}.

The variation of the gaps for the strong pairing of quarks within
the 2SC phase and the gluon-photon mass in the interval
$m_{g-\gamma}\sim 20\div 200~$MeV only slightly affects the
results. The main cooling process is the QDU process on normal
quarks. We see that the presence of normal quarks entails too fast
cooling. The data could be explained only, if all the masses lie
in a very narrow interval ($1.21<M/M_\odot<1.22$ in our case). In
case of the other two crust models the resulting picture is
similar.

The existence of only a very narrow mass interval, in which the
data can be fitted seems to us unrealistic by itself. Moreover the
observations show existence of neutron stars in binary systems
with very different masses, e.g., $M_{\rm B1913+16}\simeq 1.4408
\pm 0.0003~M_{\odot}$ and $M_{\rm J0737-3039B}\simeq 1.250 \pm
0.005~M_{\odot}$, cf. \cite{Lyne:2004cj}. {\em{Thus the data can't
be satisfactorily explained.}}

  Then we assume a possibility of an yet unknown X-pairing
channel with a small $\Delta_X$ gap and first check the case
$\Delta_X$ to be constant. For $\Delta_X \simeq 1~$MeV the cooling
is too slow \cite{Grigorian:2004jq}. This is true for all three
crust models. Thus the gaps for formerly unpaired quarks should be
still smaller in order to obtain a satisfactory description of the
cooling data.

For the $\Delta_X = 30~$ keV the cooling data can be  fitted but
have a very fragile dependence on the gravitational mass of the
configuration. Namely, we see that all data points, except the
Vela, CTA 1 and Geminga, correspond to hybrid stars with masses in
the narrow interval $M=1.21\div 1.22 ~M_\odot$

Therefore we would like to explore whether a density-dependent
X-gap could allow a description of the cooling data within a
larger interval of compact star masses.

\begin{figure}[ht]
\begin{center}
\includegraphics[height=0.65\textwidth,angle=-90]{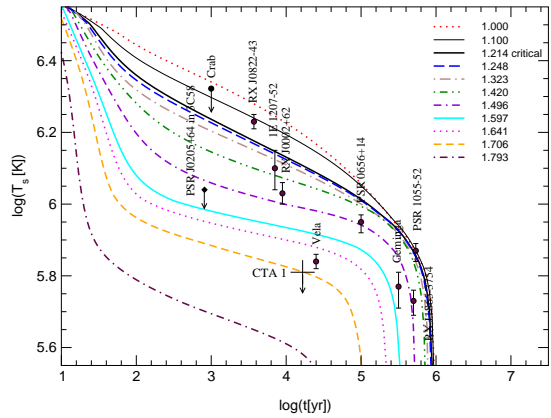}
\caption{ Cooling curves for hybrid star configurations with
Gaussian quark matter core in the 2SC phase with a density
dependent pairing gap according to Eq. (\ref{gap}) for model II. }
\label{fig:cool-csl-x-tt}
\end{center}
\end{figure}

We employ an ansatz for the X-gap as a decreasing function of the
chemical potential
\begin{equation}
\label{gap}
\Delta_X(\mu)=\Delta_c~\exp[-\alpha(\mu-\mu_c)/\mu_c]~,
\end{equation}
where the parameters are chosen such that at the critical quark
chemical potential $\mu_c=330$ MeV the onset of the
deconfinement phase transition for the X-gap has its maximal value of
$\Delta_c=1.0$ MeV and at the highest attainable chemical
potential $\mu_{\rm max}=507$ MeV, i.e. in the center of the
maximum mass hybrid star configuration, it falls to a value of the
order of $10$ keV. We choose the value $\alpha=10$ for which
$\Delta_X(\mu_{\rm max})=4.6$ keV. In Fig. \ref{fig:cool-csl-x-tt}
we show the resulting cooling curves for the gap model II with gap
ansatz eq. (\ref{gap}), which we consider as the most realistic one.

We observe that the mass interval for compact stars which obeys the
cooling data is constrained between $M=1.32~M_\odot$ for slow
coolers and $M=1.75~M_\odot$ for fast coolers such as Vela.
This results we obtain are based on a purely hadronic model with
different choices of the parameters \cite{Blaschke:2004vq}. Note that
according to a recently suggested independent test of cooling
models \cite{Popov:2004ey} by comparing results of a corresponding
population synthesis model with the Log N - Log S distribution of
nearby isolated X-ray sources the cooling model I did not pass the
test. Thereby it would be interesting to see, whether our quark
model within the gap ansatz II
could pass the  Log N - Log S test.


\section{Conclusions}

 In this lecture devoted to the modern problems of the cooling
evolution of neutron stars we have discussed different successful
scenarios aiming to explain the known temperature - age data from
observations of compact objects. Using up today known theoretical
and experimental constraints on the structure  and cooling
regulators of compact star we end up with alternative
explanations. The neutron star can be either pure hadronic or
hybrid one with a quark core in superconducting state of matter.

  Particularly, for the hybrid stars, which are more intrigued
alternative of superdense compact objects we conclude that:

\begin{itemize}
\item
Within a nonlocal, chiral quark model the critical densities for a
phase transition to color superconducting quark matter  can be low
enough for these phases to occur in compact star configurations
with masses below $1.3~M_\odot$.

\item
For the choice of the Gaussian form-factor the 2SC quark matter
phase arises at $M\simeq 1.21~M_\odot$.

\item
Without a residual pairing the 2SC quark matter phase could
describe the cooling data only if compact stars had masses in a
very narrow band around the critical mass for which the quark core
can occur.

\item

Under assumption that formally unpaired quarks can be paired with
small gaps $\Delta_X <1~$MeV  (2SC+X pairing), which values we
varied in wide limits, only for density dependent gaps the cooling
data can be appropriately fitted.
\end{itemize}

So the present day cooling data could be still explained by hybrid
stars, assuming a complex pairing pattern, where
quarks are partly strongly paired within the 2SC channel, and
partly weakly paired with gaps $\Delta_X < 1~$MeV, which are rapidly
decrease with an increase of the density.

It remains to be investigated which microscopic pairing pattern
could fulfill the constraints obtained in this work. Another
indirect check of the model could be the Log N - Log S test.

\subsection*{Acknowledgments}
The research has been supported by the Virtual Institute of the
Helmholtz Association under grant No. VH-VI-041 and by the DAAD
partnership programm between the Universities of Yerevan and
Rostock. In particular I acknowledge D. Blaschke for his active
collaboration and support. I thank my colleagues D. Blaschke, D.N.
Voskresensky, D.N. Aguilera, J. Berdermann, and  A. Reichel for
collaboration and discussions. I also thank the organizers of the
organizers of the Helmholtz International Summer School and
Workshop on {\sl Hot points in Astrophysics and Cosmology} for
their invitation to present these lectures.

------------

\end{document}